\documentclass[a4paper,conference]{IEEEtran}
\usepackage[utf8]{inputenc}
\usepackage[T1]{fontenc}
\usepackage{url}
\usepackage{ifthen}
\usepackage{cite}
\usepackage[cmex10]{amsmath}
\usepackage{ amssymb ,centernot,bbm}

\usepackage{textcomp}

\usepackage{tikz}
	\usetikzlibrary{decorations.pathreplacing,intersections}
\usepackage{subfig}
\usepackage{cite}
\usepackage{hyperref}
\usepackage{mathtools}
\title{A Novel Centralized Strategy for Coded Caching with Non-uniform Demands}

\author{%
  
   \IEEEauthorblockN{Pierre Quinton, Saeid Sahraei and Michael Gastpar}
   \IEEEauthorblockA{EPFL\\
              IPG (IC)\\
              CH-1015 Lausanne, Switzerland\\
              Email: \{pierre.quinton, saeid.sahraei, michael.gastpar\}@epfl.ch}
 
}

\begin{document}

\maketitle 

\begin{abstract}
Despite significant progress in the caching literature concerning the worst case and uniform average case regimes, the algorithms for caching  with nonuniform demands are still at a basic stage and mostly rely on simple grouping and memory-sharing techniques. In this work we introduce a novel centralized caching strategy for caching with nonuniform file popularities. Our scheme allows for assigning more cache to the files which are more likely to be requested, while maintaining the same sub-packetization for all the files. As a result, in the delivery phase it is possible to perform linear codes across files with different popularities without resorting to zero-padding or concatenation techniques. We will describe our placement strategy for arbitrary range of parameters. The delivery phase will be outlined for a small example for which we are able to show a noticeable improvement over the state of the art. \end{abstract}

\section{Introduction} 
Caching is a communication technique for redistributing the traffic in a broadcast network and thereby reducing its variability over time. The idea is to transfer part of the data to the users during low traffic periods. This data is stored at the caches of the users and helps as side information when later the server transfers the remaining data in a second phase. The central question in the caching literature is  that for a given cache size, by how much one can reduce the traffic in this second (delivery) phase, assuming that in the first (placement) phase one only  had partial or no knowledge at all of the requests of the users. There has been significant progress in answering this question under two paradigms. Firstly, when we look at the worst case delivery rate, meaning that we aim at minimizing the delivery rate for {\it any} request vector. Secondly, when we consider an average delivery rate under {\it uniform} distribution of the popularity of the files. For both of these scenarios the exact tradeoff between the size of the cache and the delivery rate has been characterized under uncoded placement \cite{maddah2014fundamental,yu2017exact} , i.e., when in the placement phase users are not permitted to perform coding across several files. 

By comparison, the question about minimizing the average delivery rate when the file popularities are non-uniform is still largely open. The main line of work \cite{niesen2017coded,zhang2015coded,ji2014average,ji2017order} consists of partitioning the files into two or more groups, where each group contains files with similar popularity. Then one performs memory-sharing between these groups: each user divides his cache into several chunks, and assigns a chunk to each group of files. Naturally, if a group includes the more popular files a larger chunk of the cache (per file) will be allocated to them. Finally in the delivery phase each group is served individually, ignoring coding opportunities between files from different groups.

This simple scheme even when restricted to two groups has been proved to be order-optimal, meaning that it achieves a rate within a constant factor of an information theoretic converse bound. Nevertheless, the fact that coding opportunities between files from different groups are ignored should be viewed as an unfortunate technical obstacle rather than a natural extension of the strategies that exist for uniform caching. The dilemma is clear: assigning unequal amounts of cache to different groups and applying the centralized caching strategy in \cite{maddah2014fundamental} for each group results in different sub-packetizations for files that belong to different groups. As a result, their sub-files will be of unequal size. It is therefore impossible to apply linear codes between different groups unless we resort to zero padding strategies or we concatenate the subfiles. Problems of the same nature - but perhaps less severe - appear if we resort to decentralized caching strategies \cite{maddah2015decentralized,niesen2017coded}. For instance, a  decentralized scheme has been proposed in \cite{ji2014average} where each user stores a set of $p_iMF$ packets of each file where $MF$ is the size of the cache measured in number of packets and $p_i$ is the probability of requesting file $i$. It is then suggested to perform coded delivery, but no practical scheme has been proposed for accomplishing this task.

Our main contribution in this paper is to propose a centralized caching strategy that bypasses this seemingly inevitable barrier. Specifically our placement strategy allows us to assign different amount of cache per file to different groups while maintaining equal sub-packetization for all the files. It is then very natural to allow for coding between files even if they do not belong to the same group. To the best of our knowledge this is the first centralized caching strategy that is specifically tailored for nonuniform file popularity. 
We will demonstrate the potential of this caching strategy by providing explicit delivery schemes for a small choice of the parameters and comparing its performance with the grouping strategies discussed earlier.

The rest of the paper is organized as follows. In Section \ref{sec:model} we will briefly describe the model. We will then move on to explaining our placement strategy in Section \ref{sec:placement}. Next, in section \ref{sec:delivery} we will describe our delivery strategy for a small choice of the parameters and compare its performance to the literature. Finally, we will conclude our work in Section \ref{sec:conclusion}.
\section{Model Description}
\label{sec:model}
\begin{figure}
\centering
\includegraphics[scale=0.9]{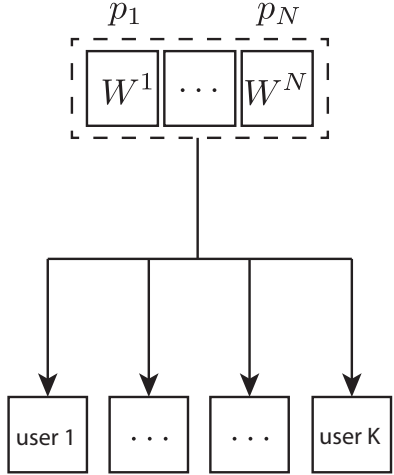}
\caption{An illustration of the caching network}
\label{fig:setup}
\end{figure}

Our model and notation will be almost identical to the one described in \cite{maddah2014fundamental}. We have a server which is in possession of $N$ independent files $\{W^1,\dots,W^N\}$ of equal size $F$ and $K$ users each equipped with a cache of size $MF$. The communication is done in two phases. In the placement phase, the server fills in the cache of each user without prior knowledge of their requests but with the knowledge of the popularity of the files. Next, each user requests precisely one file from the server. The request of each user is drawn independently from a distribution $p_{[1:N]}$ where $p_i$ represents the probability of requesting file $i$.  Note that this distribution does not vary across different users. We represent the set of requests by a vector $d$ where $d_i\in [1:N]$ for all $i\in[1:K]$. In the delivery phase the server broadcasts a message of rate $R(d)$ to satisfy all the users simultaneously. See Figure \ref{fig:setup} for an illustration. We deviate from the model in \cite{maddah2014fundamental} in that we look at the expected delivery rate instead of the peak delivery rate. We say that a memory-rate pair $(M,R)$ is achievable if and only if there exists a joint caching and delivery strategy with a cache of size $MF$ such that for any request vector $d$ a delivery message of rate $R(d)F$ satisfies all the users simultaneously, and  
\begin{eqnarray*}
R = \sum_{d}\mathbb{P}(d)R(d) = \sum_{d\in{[1:N]^K}}\prod_{i=1}^{{K}}p_{d_i}R(d).
\end{eqnarray*}

\section{The Placement Phase of Strategy $\beta$}
\label{sec:placement}
The placement phase of our strategy, which we refer to as strategy $\beta$, starts by partitioning the $N$ files into $L$ groups $G_1$,...,$G_L$ of respective size $N_1$,...,$N_L$, such that $\sum_{i=1}^L N_i = N$. How to perform this partitioning is left as a design parameter but in general files within one partition should have similar probabilities of being requested. We represent by $g_i\in[1:L]$ the group to which the file $W^i$ belongs. Accordingly, each user partitions his cache into $L$ chunks of size $M_1$,...,$M_L$ such that for any $\ell\in [1:L]$, we have $M_\ell\in \lbrace 0, N_\ell/K, 2 N_\ell/K , \dots , N_\ell \rbrace$. It should be clear that this is only possible for discrete values of $M = \sum_{i = 1}^L M_i$. The overall achievable memory-rate region will be the convex hull of all the discrete pairs $(M,R)$ which can be served by our strategy. We define 
\begin{eqnarray}r_\ell = KM_\ell/N_\ell
\label{eqn:defr}
\end{eqnarray}
 and assume without loss of generality that $r_1\ge r_2\ge \cdots\ge r_L$. Note that $r_{[1:L]}$ are integers.\\
Naturally, the following two identities hold. 
\begin{align}
\sum_{\ell=1}^{L} N_\ell r_\ell &= M K\label{eq:condition1}\\
0\leq r_\ell &\leq K&&\forall \ell \in [1:L]\label{eq:condition2}.
\end{align}
 Every file in the network regardless of which group they belong to is divided into $S$ subfiles of equal size where
\begin{align*}
S={{K}\choose{K-r_1, r_1-r_2, \dots , r_{L-1}-r_L, r_L}}.
\end{align*}
The subfiles are indexed as follows
\begin{align*}
&W^i_{\tau_1, \dots, \tau_L}&\text{where } \tau_1 &\subseteq [1:K]\\
&&\tau_j &\subseteq \tau_{j-1} &\text{ for } j&\in {2, \dots, L},\\
&&|\tau_i| &= r_i  &\text{ for } i&\in [1:L].
\end{align*}
Note that there are precisely $S$  such distinct indices.\\
For any $(i,k)$ user $k$ stores subfile $W^i_{\tau_1, \dots, \tau_L}$ in his cache if and only if $k\in \tau_{g_i}$.

At this point it may help to illustrate this placement strategy via a simple example. Let us say that we have $3$ users and $2$ files and $2$ groups such that each group contains exactly one file. Let us call the files $A= W^1$ and $B = W^2$ and assume that $r_1 = 2$ and $r_2 = 1$, so $M = \frac{r_1N_1 + r_2N_2}{K} = 1$. We must divide file $A$ into $6$ subfiles $ A =\{A_{12,1}, A_{12,2},A_{13,1},A_{13,3},A_{23,2},A_{23,3}\}$. Same division applies to file $B$. The contents of the caches of the two users are illustrated in Table \ref{tab:placement_ex}.

\begin{table}[h]
  \centering
  \begin{tabular}{|c|c|c|}
    \hline
    user 1 & user 2 & user 3 \\
    \hline
    $A_{ 12, 1}$ & $A_{ 12, 1}$ & $A_{ 13, 1}$\\
    $A_{ 12, 2}$ & $A_{ 12, 2}$ & $A_{ 13, 3}$\\
    $A_{ 13, 1}$ & $A_{ 23, 2}$ & $A_{ 23, 2}$\\
    $A_{ 13, 3}$ & $A_{ 23, 3}$ & $A_{ 23, 3}$\\
    \hline
    $B_{ 12, 1}$ & $B_{ 12, 2}$ & $B_{ 13, 3}$\\
    $B_{ 13, 1}$ & $B_{ 23, 2}$ & $B_{ 23, 3}$\\
    \hline
  \end{tabular}
  \caption{Placement phase of strategy $\beta$ for parameters $N=2$, $K=3$, $M=1$, $r_1=2$, $r_2=1$}
  \label{tab:placement_ex}
\end{table}

Let us now go back to the general placement strategy and calculate the amount of cache that user $k$ dedicates to the $\ell$'th group. By definition the index of the $k$'th user must be present in all the sets $\tau_1,\dots,\tau_\ell$ whereas its index may or may not be present in the sets $\tau_{\ell + 1},\dots,\tau_{L}$. We should divide the number of such indices $\tau_1\dots\tau_L$ by the total number of subfiles $S$ to find the amount of cache dedicated to each file in group $\ell$.
\begin{align*}
M_\ell = \frac{N_\ell }{S}& {{K-1}\choose{r_1-1}} \times \prod_{i=1}^{\ell-1}{{r_i-1}\choose{r_{i+1}-1}}\times \prod_{i=\ell}^{L-1}{{r_i}\choose{r_{i+1}}}\\
= &N_\ell \frac{{{K-1}\choose{r_1-1}}}{{{K}\choose{r_1}}} \times \prod_{i=1}^{\ell -1} \frac{{{r_i-1}\choose{r_{i+1}-1}}}{{{r_i}\choose{r_{i+1}}}}\\
= &N_\ell\frac{r_1}{K} \times\prod_{i=1}^{\ell-1} \frac{r_{i+1}}{r_i}\\
= & \frac{r_\ell N_\ell}{K}.
\end{align*}
Note that this expression matches with the way we defined the parameter $r_\ell$ in Equation \eqref{eqn:defr}. 
\section{Delivery Strategy for $K = 3, N = 2$ and Comparison To the Literature}
\label{sec:delivery}
Let us start by describing our delivery strategy for the same toy example as in the previous section. The explicit delivery messages for all possible request vectors are provided in Table \ref{tab:toy}.

\begin{table}[h]
	\centering
	\begin{tabular}{|c|c|c|}
		\hline
		request vector& delivery message &delivery rate\\
		\hline
		$( A, A, A )$ &$A_{12,1} \oplus A_{13,1}\oplus A_{23,2}$& $1/3$\\
		 &$A_{12,2} \oplus A_{13,3}\oplus A_{23,3}$& \\
		 \hline
		$( A, A, B )$ &$B_{12,1} \oplus A_{23,2}$ , $B_{13,1} \oplus A_{23,3}$& $2/3$\\
		&$B_{12,2} \oplus A_{13,1}$ , $B_{23,2} \oplus A_{13,3}$&\\
		\hline
		$( A, B, B )$ &$B_{12,1} \oplus A_{23,2}$ , $B_{13,1} \oplus A_{23,3}$ & $2/3$\\
		&$B_{12,2} \oplus B_{13,3}$ , $B_{23,2} \oplus B_{23,3}$ & \\
		\hline
		$( B, B, B )$ &$B_{12,1}\oplus B_{12,2}$ , $B_{12,1} \oplus B_{13,3}$& $2/3$\\
		&$B_{13,1}\oplus B_{23,2}$ , $B_{13,1} \oplus B_{23,3}$& \\
		\hline
	\end{tabular}
	\caption{the set of delivery messages for $N = 2, K = 3$ and $r_1 =2, r_2 = 1 $ for all possible request vectors (different permutations are omitted.)}
	\label{tab:toy}
\end{table}
Let us say that file $A$ is requested with probability $p$ and file $B$ with probability $1-p$. We assume without loss of generality that $p\ge 0.5$. The expected delivery rate is
\begin{eqnarray*}
{R} = \frac{1}{3} p^3 + \frac{2}{3}(1-p^3) = \frac{2}{3} - \frac{1}{3}p^3.
\end{eqnarray*}
Alternatively we can set $(r_1,r_2) = (3,0)$ which results in an expected delivery rate of $1-p^3$. Therefore,

\begin{eqnarray}
{R}_\beta = \min\{ \frac{2}{3} - \frac{1}{3}p^3,1-p^3\}.
\label{eqn:performance_beta}
\end{eqnarray}
Therefore, the point $(M,R) = (1,\min\{ \frac{2}{3} - \frac{1}{3}p^3,1-p^3\})$ is achievable with strategy $\beta$.
We want to compare this with the achievable rate of grouping strategy in \cite{niesen2017coded}. The strategy in \cite{niesen2017coded} is particularly designed for decentralized caching, which by nature has an inferior performance (in terms of delivery rate) compared to its centralized counterpart. Thus, before we perform the comparison we slightly modify the strategy in \cite{niesen2017coded} without compromising its basic concepts: the files are grouped in $L$ disjoint sets and each user partitions his cache into $L$ segments. Coding opportunities between several groups are ignored in the placement and delivery phase. However, instead of performing decentralized caching within each group we deploy the centralized caching strategy from \cite{maddah2014fundamental,yu2017exact}. We refer to this as strategy $\alpha$. It is easy to see that strategy $\alpha$ always outperforms the strategy in \cite{niesen2017coded} in terms of expected delivery rate. It is also easy to see that strategy $\alpha$ always performs at least as good as the strategy in \cite{maddah2014fundamental,yu2017exact} since by definition we can have only one partition which includes all the files. Let us now proceed to compare the two strategies $\alpha$ and $\beta$.

For the same choice of parameters $K=3, N=2, M = 1$, strategy $\alpha$ can be deployed with $L=1$ or $L=2$ groups. The former gives an expected rate of 
\begin{eqnarray*}
{R}_{\alpha,L=1} &=& \frac{1}{2}(p^3+ (1-p)^3) + \frac{2}{3}(1-p^3-(1-p)^3) \\
&=& \frac{2}{3} - \frac{1}{6}(p^3 + (1-p)^3).
\end{eqnarray*}
If instead we set $L =2$, we must divide the cache into two segments of sizes $M_1$ and $M_2 = 1 - M_1$. We will then ignore any coding opportunities between the files $A$ and $B$, so the delivery rate is given by
\begin{eqnarray*}
{R}_{\alpha,L=2} &=& [1-M_1]p^3 + [1-M_2] (1-p)^3 \\&+&  [(1-M_1) +(1 - M_2)] (1 - p^3 - (1-p)^3)\\
 &=& 1 - M_1p^3 - (1-M_1)(1-p)^3.
\end{eqnarray*}
Assuming $p\ge \frac{1}{2}$ it is then profitable to set $M_1 = 1$ and we get a rate of 
\begin{eqnarray*}
{R}_{\alpha,L=2} &=& 1 - p^3.
\end{eqnarray*}
To summarize, we can write
\begin{eqnarray}
{R}_\alpha =\min\{ \frac{2}{3} - \frac{1}{6}(p^3 + (1-p)^3) , 1-p^3\}.
\label{eqn:performance_alpha}
\end{eqnarray}
Comparing Equations \eqref{eqn:performance_beta}  and \eqref{eqn:performance_alpha} we see that strategy $\beta$ strictly outperforms strategy $\alpha$ as long as $\frac{1}{2}< p < (1/2)^{\frac{1}{3}}\approx 0.794$. 
Let us summarize this in a table.
{\def\arraystretch{2}\tabcolsep=10pt

\begin{table}[h]
	\centering
	\begin{tabular}{|c|c|c|}
	\hline
	probability of file A&\multicolumn{2}{|c|} {Expected Delivery Rate}\\
	\cline{2-3}
		 &strategy $\alpha$&Strategy $\beta$\\
		\hline
		$0.5 \le p \le 0.739$ &$\frac{2}{3}-\frac{1}{6}(p^3 + (1-p)^3) $& $\frac{2}{3}-\frac{1}{3}p^3$\\
		\hline
		$0.739 < p \le 0.794$ &$1-p^3 $& $\frac{2}{3}-\frac{1}{3}p^3$\\
		\hline
		$0.794 < p \le 1$ &$1-p^3 $& $1-p^3$\\
		\hline
		\end{tabular}
	\caption{Comparison of the expected delivery rate of strategies $\alpha$ and $\beta$ when $K = 3, N=2$ and $M=1$. We assume that file $A$ is requested with probability $p\ge 1/2$.}
	\label{tab:comparison}
\end{table}
}
\begin{figure}
\includegraphics[scale=0.6]{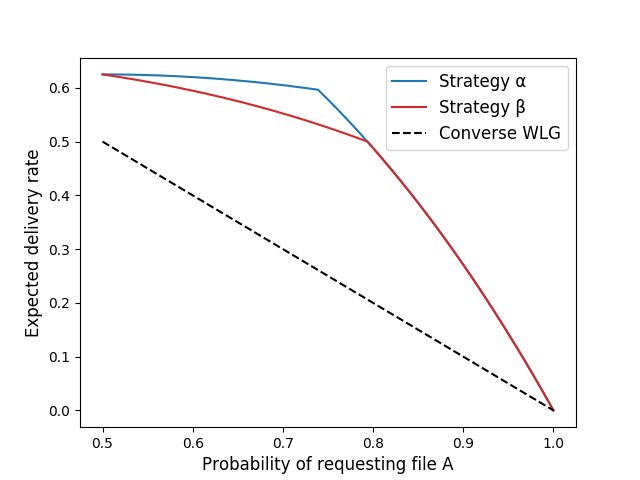}
\caption{Comparison of strategies $\alpha$ and $\beta$ resumed in Table \ref{tab:comparison} together with the converse bound from \cite{wang2016new}.}
\label{fig:versusp}
\end{figure}

In Figure \ref{fig:versusp} we compare the delivery rates of the two strategies for $N=2,K=3, M=1$. On the horizontal axis the probability of ordering file $A$ increases from $0.5$ to $1$ and on the vertical axis we have the expected delivery rate.  The maximum gain is offered over strategy $\alpha$ when $p =0.738$ in which case ${R}_\beta \approx 0.89 {R}_\alpha$. A converse bound from \cite{wang2016new} is plotted for comparison.

\begin{figure}
\includegraphics[width=0.54\textwidth]{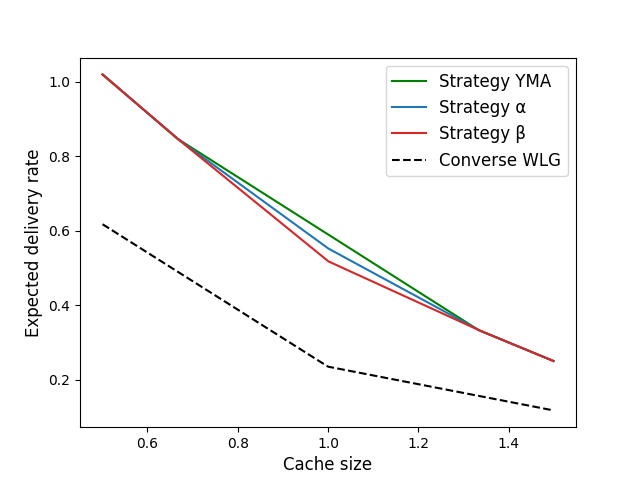}
\caption{Comparison of the expected delivery rate of strategies $\alpha$ and $\beta$ and the strategy described in \cite{yu2017exact} (YMA) when $K = 3, N=2$, together with the converse bound from \cite{wang2016new}. We assume that file $A$ is requested with probability $p=0.765$.}
\label{fig:versusM}
\end{figure}

Similar analysis can be done for other cache sizes. In Table \ref{tab:allM} we summarize the achievable rate of strategy $\beta$ for difference choices of the parameters $r_1$ and $r_2$ which results in $M = (r_1 + r_2)/K$.
{\def\arraystretch{2}\tabcolsep=10pt

\begin{table}[t]
	\centering
	\begin{tabular}{|c|c|c|}
	\hline
	 {cache size, $M$}&$(r_1,r_2)$& Expected Delivery Rate\\
		\hline
		$0$ &$(0,0) $& $2-p^3-(1-p)^3$\\
		\hline
		$\frac{1}{3}$ &$(1,0)$&$\frac{5}{3}-p^3-\frac{2}{3}(1-p)^3$\\
		\hline
		$\frac{2}{3}$ &$(1,1)$& $1- \frac{1}{3}p^3-\frac{1}{3}(1-p)^3$\\
		\hline
		$1$ &$(2,1)$& $\frac{2}{3}-\frac{1}{3}p^3$\\
		\hline		
		$1$ &$(3,0)$& $1-p^3$\\
		\hline		
		$\frac{4}{3}$ &$(2,2)$& $\frac{1}{3}$\\
		\hline
		$\frac{4}{3}$ &$(3,1)$& $\frac{2}{3}-\frac{2}{3}p^3$\\
		\hline
		$\frac{5}{3}$ &$(3,2)$& $\frac{1}{3}-\frac{1}{3}p^3$\\
		\hline
		$2$ &$(3,3)$& $0$\\
		\hline
		\end{tabular}
	\caption{The expected delivery rate of strategy $\beta$ when $K = 3, N=2$ for different values of $(r_1,r_2)$ which results in different cache sizes $M$. We assume that file $A$ is requested with probability $p\ge 1/2$.}
	\label{tab:allM}
\end{table}
}
The achievable memory-rate region for $K = 3$, $N=2$ is the convex hull of all these points. Note that depending on the value of $p$ some of these points may become irrelevant. For instance if $p = 1$, the points achieved by setting $(r_1,r_2)=(2,2)$ does not lie on the boundary of the convex hull. In Figure \ref{fig:versusM} we have plotted the achievable memory-rate region for strategies $\alpha$ and $\beta$ for $N = 2, K =3$ and for $p = 0.765$, where the improvements offered by strategy $\beta$ are most visible. Again, the converse bound from \cite{wang2016new} has been included for comparison. Note that the plot has been trimmed, since the performance is identical for very small or very large cache sizes. The gains are most visible in the vicinity of $M = 1$. 

\section{Conclusion and Future Work}
\label{sec:conclusion}

In this paper we presented a novel centralized caching strategy for non-uniform demands and demonstrated that for a small choice of parameters it outperforms the state of the art. For our future work, we intend to generalize our delivery strategy to arbitrary range of parameters. It is noteworthy that our strategy has the potential to be adapted to a user-specific popularity scenario, that is when the probability of requesting different files varies across the users. This can serve as another interesting direction for future research.

\bibliographystyle{IEEEtran}
\bibliography{IEEEfull,nonuniform}

\end{document}